%% file: conference_paper.tex
\documentclass[conference]{IEEEtran}
\usepackage{cite}
\usepackage{amsmath,amssymb,bm}
\usepackage{graphicx}
\usepackage{booktabs}
\usepackage{url}
\usepackage{eso-pic}
\usepackage[nospread,noshrink,checkfloat]{flushend}
\newcommand{\E}{\mathbb{E}}
\newcommand{\Prb}{\mathbb{P}}
\newcommand{\CN}{\mathcal{CN}}
\newcommand{\INR}{\mathsf{INR}}
\newcommand{\vect}[1]{\bm{#1}}
\makeatletter
\newcommand{\resultmark}{%
  \def\tagform@##1{%
    \maketag@@@{\shortstack[t]{(\ignorespaces##1\unskip\@@italiccorr)\\[-0.3ex]$\square$}}}}
\makeatother
\DeclareMathOperator{\erfc}{erfc}

\DeclareMathOperator*{\argmin}{arg\,min}

\begin{document}
\AddToShipoutPictureFG*{%
  \AtPageUpperLeft{%
    \raisebox{-0.45in}[0pt][0pt]{%
      \makebox[\paperwidth]{%
        \parbox[b]{\textwidth}{\centering\footnotesize
          This work has been submitted to the IEEE for possible publication.\\
          Copyright may be transferred without notice, after which this version may no longer be accessible.}}}}}
\title{Delay--Doppler Synchronization for Distributed\\
Uplink Interference Nulling in LEO Networks}
\input{paper_sector/authors}
\maketitle
\suppressfloats[t]
\bstctlcite{BSTcontrol}  

\input{paper_sector/conference_short/abstract}
\input{paper_sector/conference_short/intro}
\input{paper_sector/conference_short/method}
\input{paper_sector/conference_short/numerical}
\input{paper_sector/conference_short/long_term_evaluation}
\input{paper_sector/conference_short/conclusion}


\bibliographystyle{IEEEtran}
\bibliography{references}

\end{document}

%% file: paper_sector/authors.tex
\author{%
  \IEEEauthorblockN{Stefan Daniel Tancovici\IEEEauthorrefmark{1},
    Shizhen Jia\IEEEauthorrefmark{3}, \\Marco Mezzavilla\IEEEauthorrefmark{1},
    C. Nicolas Barati\IEEEauthorrefmark{2}, and Sundeep Rangan\IEEEauthorrefmark{3}}
  \IEEEauthorblockA{%
    \IEEEauthorrefmark{1}Politecnico di Milano, Milan, Italy\\
    \IEEEauthorrefmark{2}Old Dominion University, Norfolk, VA, USA\\
    \IEEEauthorrefmark{3}NYU WIRELESS, New York University, Brooklyn, NY, USA\\[0.3ex]
    {\small stefandaniel.tancovici@mail.polimi.it, marco.mezzavilla@polimi.it}\\[-0.3ex]
    {\small cbaratin@odu.edu, \{s.jia, srangan\}@nyu.edu}}%
}

%% file: paper_sector/conference_short/abstract.tex
\begin{abstract}
Cooperation among low-Earth-orbit (LEO) satellites provides spatial
diversity for mitigating uplink interference that is difficult to
suppress with a single onboard array. Coherent interference cancellation
across satellites, however, requires compensation of the relative delay
and Doppler of a common interferer. We analyze
two-branch interference nulling with finite training and derive a
closed-form approximation to the mean residual interference-to-noise
ratio (INR). The approximation separates wrong acquisition, local
synchronization errors, and weight-learning losses, linking performance
to input INR, branch power imbalance, training length, and effective
search size. Under a Gaussian surrogate with small errors and efficient,
approximately unbiased refinement, correct acquisition yields a high-INR
plateau of $2/N_v$, where $N_v$ is the valid training length.
Synchronization and weight learning contribute equally to this limit,
whereas persistent alignment bias adds a penalty that grows with input
power. Monte Carlo experiments compare dense FFT/zoom and coarse
FFT/gradient acquisition using matched final FIR refinement, branch-noise
normalization, and independent test blocks. Both approach the predicted
plateau at high input INR. Coarse search reduces measured search time
by $6.7\times$ at an approximately 0.7-dB acquisition-threshold penalty.
Controlled power-ratio sweeps show how branch imbalance delays reliable
acquisition at fixed stronger-branch INR. A separate Starlink snapshot
evaluation demonstrates comparable nulling performance under changing
satellite geometry.
\end{abstract}
\begin{IEEEkeywords}
Satellite communications, interference nulling, delay--Doppler estimation,
sample covariance, residual INR.
\end{IEEEkeywords}

%% file: paper_sector/conference_short/intro.tex
\section{Introduction}
\label{sec:introduction}
\ifdefined\input{paper_sector/conference_short/network_scenario}

\input{paper_sector/conference_short/network_scenario}
\fi
Low-Earth-orbit (LEO) constellations extend wireless
coverage~\cite{kodheli2021satellite,rinaldi2020ntn}, but their uplinks
face ground-based interference and jamming~\cite{yue2023leo}.
Onboard beamforming guided by radio environment
maps~\cite{caceres2024interference} and null shaping under location
uncertainty~\cite{caceres2025nullshaping} mitigate interference.
However, small angular separation between a nearby
jammer and the desired transmitter can limit interference rejection by
a single onboard array.

Our previous work~\cite{jia2026distributed} studied concurrent uplink
transmission from a ground terminal to multiple spatially separated
satellites, exploiting channel differences for spatial diversity.
A designated master satellite can collect their received baseband
streams via inter-satellite links for alignment and interference nulling.
The prior analysis demonstrated achievable-rate gains through
game-theoretic beamforming under per-satellite equalization, leaving
distributed synchronization unresolved.

Coherent nulling requires aligning the common interferer's replicas
despite relative delay and Doppler from different path lengths and
velocities. Desired-user synchronization generally leaves these offsets
unresolved. We quantify their impact on
finite-training nulling in a two-satellite receiver.
\ifdefined\input{paper_sector/conference_short/network_scenario}
Figure~\ref{fig:network_scenario} illustrates this two-branch setting.
\fi

Prior work addresses timing/Doppler mismatch in distributed downlink
beamforming~\cite{xu2024enhancement}, terminal-side compensation for
massive-MIMO transmission~\cite{you2020massive}, and delay-robust
cooperative downlink beamforming and timing advance~\cite{chen2024asynchronous}.
These designs synchronize
communication links, leaving alignment of a separate ground interferer
unresolved.

Generalized cross-correlation estimates delay between noisy
replicas~\cite{knapp1976generalized}. Cross-ambiguity processing adds
Doppler estimation~\cite{stein1981algorithms}. The Keystone transform
reduces wideband estimation cost~\cite{xiao2018lowcomplexity}.
These methods target parameter estimation, without analyzing subsequent
finite-training nulling. Cram\'er--Rao bounds for unknown waveforms
characterize local estimation precision~\cite{chen2019unknown}, but do
not quantify wrong-peak acquisition. Delay-error bounds capture
ambiguity-driven threshold effects~\cite{weiss1983fundamental}, without
propagating them to cancellation performance.

Spatial filtering~\cite{vanveen1988beamforming} and classical
finite-sample adaptive-array analysis~\cite{reed1974rapid} characterize
beamforming and weight-learning losses, without jointly modeling
delay--Doppler acquisition. Covariance-based coexistence
nulling~\cite{jia2026noncoherent} relates spatial estimation to residual
leakage, but uses local array snapshots without distributed waveform
alignment. The remaining gap is a mean residual interference-to-noise
ratio (INR) analysis combining wrong acquisition, local synchronization
errors, and finite-training weight learning.

We study a common interferer and thermal noise at two receive branches.
Desired-user
timing/carrier synchronization, channel estimation, and signal-preserving
combining remain future work, with equalization needed when intersymbol
interference is present~\cite{tse2005fundamentals}. Residual INR alone
does not establish desired-signal preservation or achievable rate.

\noindent\textbf{Our contributions are as follows:}
\begin{itemize}
\setlength{\itemsep}{1pt}
\item \textbf{Closed-form residual-INR analysis:} We derive an
approximation separating wrong acquisition, local synchronization, and
finite-training weight-learning losses. It connects input INR, branch
power ratio, training length, and effective search size, predicts the
$2/N_v$ high-INR plateau under stated assumptions, and quantifies
fixed-bias penalties.
\item \textbf{Receiver comparison and acquisition--cost tradeoff:}
We compare dense FFT/zoom and coarse FFT/gradient acquisition with
matched final FIR refinement, noise-normalized covariance nulling, and
independent test blocks. Coarse search reduces measured search time
while retaining comparable high-INR nulling, at a higher acquisition
threshold.
\item \textbf{Power-imbalance and constellation evaluation:}
Controlled INR and power-ratio sweeps characterize the weaker branch's
effect on acquisition and assess the approximation's accuracy. A separate
24-hour Starlink snapshot study evaluates nulling under changing geometry
and specified satellite-pairing constraints.
\end{itemize}

%% file: paper_sector/conference_short/network_scenario.tex
\begin{figure}[t]
\centering
\includegraphics[width=\columnwidth]{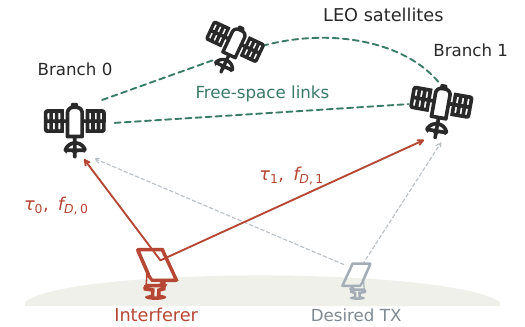}
\par\vspace{-3pt}
\caption{Cooperative LEO interference reception. Red paths carry the
common interferer. Green links enable cooperation. Pale desired uplinks
provide context. The model includes interference and noise.}
\label{fig:network_scenario}
\vspace{-3pt}
\end{figure}

%% file: paper_sector/conference_short/method.tex
\section{System Model and Receiver}
\label{sec:system}
\subsection{Two Observations of a Common Interferer}
The two satellite branches in Fig.~\ref{fig:network_scenario} observe
one common interferer and independent thermal noise. A unit-norm
maximum-ratio combiner at an $L$-element array with single-path channel
$\vect a_k=\alpha_k\vect s_k$, $\|\vect s_k\|^2=L$, gives scalar gain
$|g_k|^2=L|\alpha_k|^2$ while preserving element-noise power $N_0$.
Define the input INRs and power ratio by
\begin{equation}
 \gamma_k=\frac{P_{\rm tx}|g_k|^2}{N_0},\qquad
 \eta=\frac{\gamma_1}{\gamma_0}\leq1,\qquad
 \Gamma=\gamma_0+\gamma_1.
 \label{eq:power}
\end{equation}
After noise normalization, the observations are
\begin{equation}
 y_k[n]=\sqrt{\gamma_k}e^{j\theta_k+j2\pi f_k n}
 x[n-\bar\tau_k]+d_k[n],\quad k\in\{0,1\},
 \label{eq:streams}
\end{equation}
where $x$ has unit average power, $d_k[n]\sim\CN(0,1)$,
$\theta_0=0$, and $\theta_1$ is an unknown constant phase. Fractional
arguments denote interpolation. For sample interval $T_{\rm w}$,
$\bar\tau_k=\tau_k/T_{\rm w}$ and $f_k=f_{D,k}T_{\rm w}$. The relative
parameters are $\Delta=\bar\tau_1-\bar\tau_0$ samples and
$\nu=f_1-f_0$ cycles/sample ($N\nu$ frequency bins for $N$ samples).
Both absolute delays are retained in simulation because the shaped QPSK
waveform is cyclostationary. The analysis uses a stationary surrogate.

\subsection{Acquisition, Alignment, and Nulling}
\begin{figure}[!t]
\centering
\includegraphics[width=0.9\columnwidth]{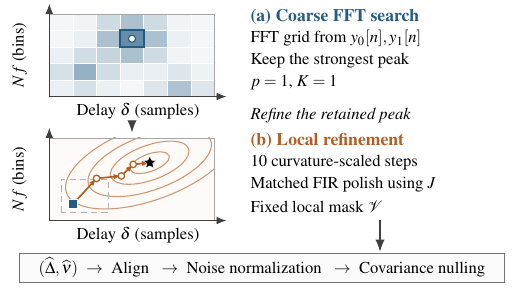}
\par\vspace{-3pt}
\caption{Coarse acquisition and refinement ($p=1$, $K=1$), followed by
alignment, branch-noise normalization, and nulling. The landscape and
iterates are schematic.}
\label{fig:method_overview}
\vspace{-3pt}
\end{figure}
For a delay--Doppler candidate $(\delta,f)$, compensate branch 1 as
\begin{equation}
 \widetilde y_1[n;\delta,f]=e^{-j2\pi f(n+\delta)}y_1[n+\delta].
 \label{eq:alignment}
\end{equation}
At $(\Delta,\nu)$, the two interference copies differ only by a constant
complex gain, absorbed by the nuller. An unnormalized FFT cross-ambiguity
search initializes either dense zoom or coarse gradient refinement.
Both then maximize the same normalized score over a local box
$\mathcal B$:
\begin{equation}
 J(\delta,f)=
 \frac{\left|\sum_{n\in\mathcal V}y_0[n]\widetilde y_1^*[n;\delta,f]\right|^2}
 {\sum_{n\in\mathcal V}|y_0[n]|^2\sum_{n\in\mathcal V}|\widetilde y_1[n;\delta,f]|^2}.
 \label{eq:score}
\end{equation}
The valid mask $\mathcal V$ excludes interpolation transients and
nonoverlap and is fixed within each local box. The same de-rotation/FIR
operator is used for refinement and nulling. A bounded L-BFGS-B solver
approximately maximizes $J$. Write $N_v=|\mathcal V|$.
Figure~\ref{fig:method_overview} illustrates the coarse path: the
strongest FFT peak seeds curvature-scaled gradient updates, followed by
matched final FIR refinement. A local numerical
maximum does not guarantee statistical efficiency. The analysis treats
the overlap length as a common $N_v$, although overlap can depend on
the selected delay.

Using the selected alignment, form $\vect z[n]=[y_0[n],\widetilde
 y_1[n]]^T$ and $\widehat{\vect R}=N_v^{-1}\sum_{n\in\mathcal V}
\vect z[n]\vect z[n]^H$. For aligned noise covariance $\vect R_d$,
\begin{equation}
 \widehat{\vect w}=\argmin_{\vect w\ne0}
 \frac{\vect w^H\widehat{\vect R}\vect w}{\vect w^H\vect R_d\vect w},
 \qquad \widehat{\vect w}^H\vect R_d\widehat{\vect w}=1.
 \label{eq:nuller}
\end{equation}
This is the minimum-eigenvalue rule after branch-noise whitening.
For FIR coefficients $h_\tau[\ell]$ and temporally white input noise,
$\vect R_d=\operatorname{diag}(1,\sum_\ell|h_\tau[\ell]|^2)$. This
normalizes branch powers without whitening temporal correlation.
With interference covariance $\vect S$, the target metric is
\begin{equation}
 \overline{\INR}=\E_{\rm train}
 [\widehat{\vect w}^H\vect S\widehat{\vect w}].
 \label{eq:metric}
\end{equation}
Independent test blocks estimate this residual, avoiding the optimistic
training minimum eigenvalue. No desired signal enters the covariance.

\section{Residual Interference Analysis}
\label{sec:theory}
Let $\mathcal C$ denote correct main-lobe acquisition and
$P_w=\Prb(\mathcal C^c)$. Conditioning gives the exact identity
\begin{equation}
 \overline{\INR}=P_w\overline{\INR}_w+(1-P_w)\overline{\INR}_c.
 \label{eq:mixture}
\end{equation}
We approximate its terms using equal branch-noise powers, Gaussian
surrogate statistics, effective independence of training samples and
search cells, small synchronization errors, and efficient, approximately
unbiased refinement. Sample correlation, selected-cell conditioning,
and training-data reuse limit the approximation.

\subsection{Correct Acquisition: Two Residual Contributions}
With aligned waveform coherence $r$, the interference covariance is
\begin{equation}
 \vect S=\begin{bmatrix}
 \gamma_0&\sqrt{\gamma_0\gamma_1}r^*\\
 \sqrt{\gamma_0\gamma_1}r&\gamma_1
 \end{bmatrix}.
\end{equation}
Its eigenvalues are
\begin{equation}
 s_\pm=\frac{\Gamma\pm
 \sqrt{\Gamma^2-4\gamma_0\gamma_1(1-|r|^2)}}{2}.
 \label{eq:eigen}
\end{equation}
Perfect alignment makes $\vect S$ rank one. Expanding its smaller
eigenvalue near $|r|=1$ gives the population alignment loss
\begin{equation}
 s_-\simeq\frac{2\gamma_0\gamma_1}{\Gamma}(1-|r|).
 \label{eq:coherence_cost}
\end{equation}
The population covariance $\vect R=\vect I+\vect S$ has eigenpairs
$(\lambda_\pm,\vect v_\pm)$, with $\lambda_\pm=1+s_\pm$.
Finite training perturbs its minimum-eigenvalue eigenvector toward the
dominant mode. Writing $\vect E=\widehat{\vect R}-\vect R$,
first-order perturbation gives
\begin{equation}
 \widehat{\vect w}\simeq
 \frac{\vect v_-+\epsilon\vect v_+}{\sqrt{1+|\epsilon|^2}},\qquad
 \epsilon=\frac{\vect v_+^H\vect E\vect v_-}{\lambda_--\lambda_+}.
 \label{eq:eigenvector_perturbation}
\end{equation}
Substitution into the independent-test quadratic form yields
\begin{equation}
 \widehat{\vect w}^H\vect S\widehat{\vect w}
 \simeq\frac{s_-+s_+|\epsilon|^2}{1+|\epsilon|^2}
 \simeq s_-+(s_+-s_-)|\epsilon|^2.
 \label{eq:residual_perturbation}
\end{equation}
Independent Gaussian samples give
$\E|\epsilon|^2=\lambda_+\lambda_-/[N_v(\lambda_+-\lambda_-)^2]$.
Since $s_+-s_-=\lambda_+-\lambda_-$, the additional interference is
\begin{equation}
 \INR_{\rm learn}\simeq
 \frac{\lambda_+\lambda_-}{N_v(\lambda_+-\lambda_-)}
 \simeq\frac{1+\Gamma}{N_v\Gamma},
 \label{eq:learning}
\end{equation}
requiring $N_v\Gamma^2/(1+\Gamma)\gg1$.
Weight selection biases the training minimum eigenvalue downward.
Equation~\eqref{eq:learning} instead measures learned-weight leakage on independent
tests, as required by \eqref{eq:metric}.

For a contiguous valid window, the stationary, separable model gives
\begin{equation}
 |r|\simeq|R_x(\delta\tau)|
 \left|\frac1{N_v}\sum_{n=0}^{N_v-1}
 e^{j2\pi\delta\nu(n-\bar n)}\right|.
 \label{eq:coherence_factorization}
\end{equation}
Here $R_x(0)=1$ and $\bar n=(N_v-1)/2$. Constant phase drops out.
Expanding both factors for small errors $(\delta\tau,\delta\nu)$ gives
\begin{equation}
 1-|r|\simeq\frac{(2\pi)^2}{2}
 (\sigma_f^2\delta\tau^2+V_t\delta\nu^2),\quad
 V_t=\frac{N_v^2-1}{12},
 \label{eq:coherence}
\end{equation}
where $\sigma_f^2$ is normalized spectral variance and $V_t$ is the
variance of the centered sample indices. Combining
\eqref{eq:coherence_cost}, \eqref{eq:learning}, and \eqref{eq:coherence}
gives a useful form before imposing efficient estimation:
\begin{equation}
 \overline{\INR}_c\simeq\frac{1+\Gamma}{N_v\Gamma}
 +\frac{\gamma_0\gamma_1}{\Gamma}(2\pi)^2
 (\sigma_f^2 m_\tau+V_t m_\nu),
 \label{eq:residual_mse}
\end{equation}
where $m_\tau=\E[\delta\tau^2\mid\mathcal C]$ and
$m_\nu=\E[\delta\nu^2\mid\mathcal C]$ include variance and squared
bias. This form quantifies synchronization loss without assuming
statistical efficiency. Define
$C=\gamma_0\gamma_1/[(1+\gamma_0)(1+\gamma_1)]$.
A centered, separable surrogate information model uses
\begin{equation}
 I_\tau\simeq N_v(2\pi)^2\sigma_f^2\frac{2C}{1-C},\qquad
 I_\nu\simeq N_v(2\pi)^2V_t\frac{2C}{1-C}.
 \label{eq:information}
\end{equation}
Centering the time index accounts for the unknown constant phase in this
model. Assuming error variances $I_\tau^{-1}$ and $I_\nu^{-1}$, each
parameter contributes $(1-C)/(4CN_v)$ to $\E[1-|r|]$. Since
$\gamma_0\gamma_1(1-C)/C=1+\Gamma$, substitution into
\eqref{eq:coherence_cost} and addition of \eqref{eq:learning} yield
\begin{equation}
 \resultmark
 \E[s_-]\simeq\frac{1+\Gamma}{N_v\Gamma},\qquad
 \overline{\INR}_c\simeq\frac{2(1+\Gamma)}{N_v\Gamma}.
 \label{eq:correct}
\end{equation}
Alignment and weight learning each contribute half, giving the
high-INR limit $2/N_v$. Equation~\eqref{eq:information} is not an exact
finite-window CRB for correlated RRC-shaped QPSK. The approximation
cannot be extrapolated to vanishing power or biased refinement.
Cancellation of spectral and temporal variance factors requires both
parameters to attain their surrogate information scales. Colored samples
and training-data reuse can alter the penalties and their interaction.

\subsection{Incorrect Acquisition and the Mean Residual}
At a wrong cell, decorrelated replicas give
$\vect S\simeq\operatorname{diag}(\gamma_0,\gamma_1)$. Put
$a=1+\gamma_0$, $b=1+\gamma_1$. Approximate the unnormalized correlation
by $s\sim\CN(0,N_vab)$, so $u=|s|^2/(N_vab)\sim\operatorname{Exp}(1)$.
Their maximum $U_{\max}$ over $M$ independent cells satisfies, for
integer $M$ and $u\geq0$,
\begin{equation}
 \begin{split}
 F_{U_{\max}}(u)&=(1-e^{-u})^M,\\
 \E[U_{\max}]&=\sum_{m=1}^{M}\frac1m
 \simeq\ln M+\gamma_{\rm E}\equiv q_M,
 \end{split}
 \label{eq:maximum_statistics}
\end{equation}
with Euler's constant $\gamma_{\rm E}\simeq0.577$. We use the smooth
approximation $q_M$ for an effective, possibly noninteger cell count.
Replacing the selected off-diagonal covariance
power by $abq_M/N_v$ and neglecting diagonal fluctuations gives
\begin{align}
 \cos\vartheta&=\frac{\gamma_0-\gamma_1}
 {\sqrt{(\gamma_0-\gamma_1)^2+4abq_M/N_v}},\label{eq:costheta}\\
 \overline{\INR}_w&\simeq\gamma_1+
 (\gamma_0-\gamma_1)\frac{1-\cos\vartheta}{2}.\label{eq:wrong}
\end{align}
The selected minimum eigenvector has branch powers
$|\widehat w_0|^2\simeq(1-\cos\vartheta)/2$ and
$|\widehat w_1|^2\simeq(1+\cos\vartheta)/2$. Weighting the independent
test powers by these quantities gives \eqref{eq:wrong}. The residual
approaches $\gamma_1$ for weak mixing ($\cos\vartheta\to1$) and $\Gamma/2$ for strong mixing ($\cos\vartheta\to0$).

At the correct cell, the surrogate
$s\sim\CN(N_v\sqrt{\gamma_0\gamma_1},N_vab)$ implies
$2u\sim\chi'^2_2(2N_vC)$. Approximating
$\sqrt u\simeq\sqrt{N_vC}+\mathcal N(0,1/2)$ and comparing with
$\sqrt{q_M}$ yields
\begin{equation}
 P_w\simeq\tfrac12\erfc\bigl(\sqrt{N_vC}-\sqrt{q_M}\bigr).
 \label{eq:pw}
\end{equation}
Here $M$ counts effective independent cells, not oversampled grid points.
The replacement of a random maximum by $q_M$ omits its fluctuations and
inter-cell dependence. Within this approximation, the $P_w=1/2$
transition satisfies $N_vC=q_M$. If both branch INRs are small,
$C\simeq\eta\gamma_0^2$, so
$\gamma_{0,\mathrm{tr}}\simeq\sqrt{q_M/(N_v\eta)}$.
Thus longer training improves acquisition, a larger search region raises
the threshold logarithmically through $q_M$, and power imbalance delays
reliable synchronization even at unchanged stronger-branch INR.
Combining \eqref{eq:correct}, \eqref{eq:wrong}, and \eqref{eq:pw} gives
\begin{equation}
 \resultmark
 \begin{split}
 \overline{\INR}(\gamma_0)\simeq{}&P_w\left[\gamma_1+
 (\gamma_0-\gamma_1)\frac{1-\cos\vartheta}{2}\right]\\
 &+(1-P_w)\frac{2(1+\Gamma)}{N_v\Gamma}.
 \end{split}
 \label{eq:inr_final}
\end{equation}
With $\gamma_1=\eta\gamma_0$, the expression connects input INR,
power imbalance, training length, and search size. It predicts an
acquisition transition followed by the $2/N_v$ plateau, subject to the
stated surrogate assumptions.

\input{paper_sector/conference_short/results_figure}

\subsection{Fixed Bias and Coarse Acquisition}
Under the small-error model in \eqref{eq:residual_mse}, fixed delay and
Doppler biases $(b_\tau,b_\nu)$ add
\begin{equation}
 \Delta\INR_{\rm bias}\simeq\frac{\gamma_0\gamma_1}{\Gamma}(2\pi)^2
 (\sigma_f^2b_\tau^2+V_tb_\nu^2).
 \label{eq:bias}
\end{equation}
At fixed $\eta$ and nonzero biases, this penalty grows linearly with
input power and can exceed the nominal $2/N_v$ plateau.

Coarse sampling of the ambiguity surface can lower the sampled
main-lobe peak and increase the risk of wrong acquisition. After correct
acquisition, matched final FIR refinement can reduce grid-quantization
errors. Identical refinement settings do not guarantee equal estimation
accuracy or comparable wrong-peak statistics.

%% file: paper_sector/conference_short/results_figure.tex
\begin{figure*}[!t]
\centering
\includegraphics[width=\textwidth]{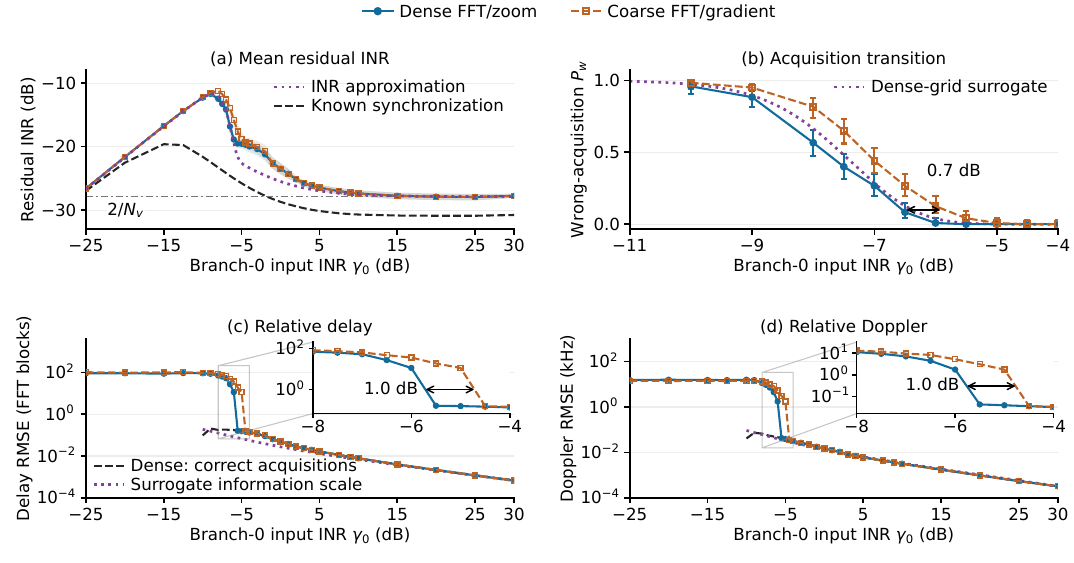}
\par\vspace{-3pt}
\caption{Dense and coarse acquisition with matched FIR refinement:
(a) mean residual INR and pointwise 95\% bootstrap intervals,
(b) wrong-acquisition rates and Wilson intervals,
(c), (d) delay and Doppler RMSE, with transition zooms and equal-RMSE
gap arrows. Black dashed: known synchronization in (a), correct dense
acquisitions in (c), (d). Purple dotted: surrogate references.
In (a), the dotted curve is \eqref{eq:inr_final}, which reduces to
\eqref{eq:correct} when acquisition failures are negligible. The gray
horizontal line marks the high-INR limit $2/N_v$ of \eqref{eq:correct}.
The information scales in (c), (d) are not exact finite-window CRBs.}
\label{fig:results}
\vspace{-3pt}
\end{figure*}

%% file: paper_sector/conference_short/numerical.tex
\section{Numerical Results}
\label{sec:numerical}
\input{paper_sector/conference_short/validation_figures}
\subsection{Configuration and Matched Receivers}
Sionna RT 2.2~\cite{sionna} regenerates two line-of-sight array channels
from the first archived Starlink record~\cite{celestrak}. A fixed archived
unit-norm beam transmits RRC-shaped QPSK (Table~\ref{tab:parameters}).
The sample rate is 585.9375~kHz and occupied bandwidth is 244.14~kHz.
The nominal 200-MHz link-budget bandwidth sets the original noise power
and $T_{\rm FFT}=5.12~\mu$s, not the probe bandwidth. Absolute delays and
symbol-clock phase are retained in 3816-sample periodic records with
1140-sample guards.

The effective cell count is fixed by the archived search region and
waveform bandwidth. With delay limit $D$, normalized Doppler limit
$\nu_{\max}$, and normalized occupied bandwidth $B_x=(1+\beta)/S$,
the time--bandwidth estimate is
\begin{equation}
 M\simeq(2D+1)B_x(2\nu_{\max}N+1).
 \label{eq:effective_cells}
\end{equation}
The delay and Doppler cell widths are approximately $1/B_x$ samples
and $1/N$ cycles/sample. The tabulated ranges give $M\simeq5.30\times10^4$.
Zero padding increases grid density, not the physical cell count.
Cell correlation limits the independence approximation.

\begin{table}[!t]
\vspace{-2pt}
\caption{Fixed-geometry configuration (rounded where shown).}
\label{tab:parameters}
\centering\footnotesize
\setlength{\tabcolsep}{4pt}
\begin{tabular}{@{}ll@{}}
\toprule
Parameter & Value\\
\midrule
Elements per array & $6\times6$\\
Sample interval $T_{\rm w}$ & $1.7067~\mu$s\\
Record / valid samples & $N=1536$ / $N_v=1222$\\
Power ratio / INRs at 50-dBm TX & $0.380$ / $8.98$, $4.78$ dB\\
Absolute delays (samples) & $759.607$ / $1026.077$\\
Relative Doppler & $-7.876$ kHz\\
RRC roll-off / samples per symbol & $\beta=0.25$ / $S=3$\\
FIR interpolator & 96-tap Kaiser sinc, shape 9\\
Delay / Doppler search range & $\pm600$ samples / $\pm20$ kHz\\
Training trials / test blocks per trial & $100$--$300$ / $32$\\
\bottomrule
\end{tabular}
\vspace{-3pt}
\end{table}

The input sweep has 31 settings from $-25$ to $30$~dB, sampled more
finely near acquisition. Receivers share training/test waveforms within
trials. Resetting seed 95000 at each power provides common random
numbers across powers. Residuals are averaged linearly over independent
test blocks and training trials before conversion to dB. Pointwise 95\%
intervals use 2000 bootstrap resamples of trial means. Acquisition
intervals use Wilson's method. Correct acquisition requires
$|\widehat\Delta-\Delta|\leq S/2$ and $N|\widehat\nu-\nu|\leq0.5$.

Table~\ref{tab:search} summarizes acquisition settings. Both receivers
then apply bounded L-BFGS-B refinement in a $\pm0.6$-sample,
$\pm0.6$-bin box, clipped to search limits, with at most 30 iterations.
Updates must not decrease $J$. They share the FIR operator, fixed local
mask, and noise normalization. The oracle knows the synchronization
parameters but still learns weights from finite training.

\subsection{Acquisition, Residual INR, and Cost}
Figure~\ref{fig:results} separates acquisition failure from local
estimation loss. At $\gamma_0=-7$~dB, dense and coarse search have
wrong-acquisition rates of 26.7\% and 44.2\%, and residual INRs of
$-14.17$ and $-12.61$~dB, respectively. Coarse search reaches $P_w=0.1$
about 0.7~dB later, interpolating the measured curves. Wilson intervals
remain nonzero even at zero observed errors (upper endpoint 3.1\% for
120 trials).

\begin{table}[!t]
\vspace{-2pt}
\caption{Acquisition settings and median search/refinement time.}
\label{tab:search}
\centering\footnotesize
\setlength{\tabcolsep}{4pt}
\begin{tabular}{@{}llll@{}}
\toprule
Search & FFT $p$ & Initial refinement & \textbf{Time (ms)}\\
\midrule
Dense & 4 & 6 zoom stages & \textbf{99.0}\\
Coarse & 1 & 10 gradient steps & \textbf{14.7}\\
\bottomrule
\end{tabular}
\par\vspace{2pt}
\raggedright Coarse: $K=1$ peak, gradient step factor $\mu=1$.
\vspace{-3pt}
\end{table}

After reliable acquisition, both methods give approximately $-27.25$~dB
residual INR at 8.98-dB input and $-27.75$~dB at 30-dB input, close to
$10\log_{10}(2/1222)=-27.86$~dB. At $-5$~dB input, however, the dense
residual is $-19.55$~dB versus the predicted $-22.59$~dB. Correlated,
cyclostationary samples, finite-window selection, and imperfect local
efficiency limit the surrogate's accuracy.

The RMSE panels show wrong-peak errors at low input INR and convergence
toward the common information scale after reliable acquisition. Delay
errors are divided by $S=3$ for FFT-block units. Doppler errors are
converted to kHz using $T_{\rm w}$. Equation~\eqref{eq:information} is a
model-dependent reference, not a strict bound for this QPSK experiment
or for RMSE conditioned on successful acquisition.

On one Intel Core Ultra 9 275HX thread, 40 runs after two warmups at
8.98-dB input measure FFT search, candidate extraction, and all
refinement (Table~\ref{tab:search}), excluding channel generation,
weight learning, and testing. Coarse search is $6.7\times$ faster.

\subsection{Controlled Power Imbalance}
\label{sec:eta_sweep}
Figure~\ref{fig:eta_sweep} varies $\eta$ over 13 logarithmic points
from $10^{-3}$ to 1 plus the original ratio $\eta_0\simeq0.380$, for
$\gamma_0\in\{-5,0,5,10\}$~dB. Scaling branch-1 gain by
$\sqrt{\eta/\eta_0}$ preserves geometry, phases, delays, Dopplers, and
noise. Each new point uses 120 paired trials and 32 independent tests.
Original-ratio anchors reuse 100--120 trials. The additional FFT searches
are GPU-accelerated and checked against the CPU code. Table~\ref{tab:search}
retains the original CPU benchmark.

A stronger weak branch improves acquisition through $C$, but also
increases the residual available after wrong acquisition. Thus a small
residual at very small $\eta$ need not imply successful synchronization.
At $\gamma_0=0$~dB and $\eta=0.03162$, dense/coarse methods have 8/120
and 26/120 wrong acquisitions and residuals of $-20.75$/$-18.18$~dB.
At 10-dB input and $\eta=1$, both give $-27.65$~dB, matching the
surrogate. At $-5$~dB and $\eta=1$, neither has an observed acquisition
error, yet the dense residual is $-21.04$~dB versus $-23.74$~dB predicted:
reliable acquisition alone does not ensure efficient refinement.

%% file: paper_sector/conference_short/validation_figures.tex
\begin{figure}[!t]
\centering
\includegraphics[width=\columnwidth]{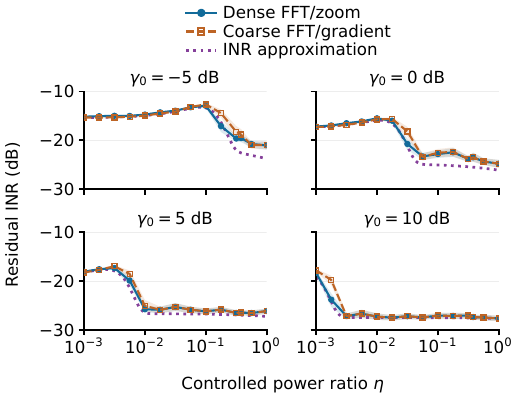}
\par\vspace{-3pt}
\caption{Residual INR versus power ratio at four fixed input INRs.
Shading: pointwise 95\% trial-bootstrap intervals. Dotted curves:
\eqref{eq:inr_final} with $N_v=1222$.}
\label{fig:eta_sweep}
\vspace{-3pt}
\end{figure}

%% file: paper_sector/conference_short/long_term_evaluation.tex
\begin{figure}[!t]
\centering
\includegraphics[width=\columnwidth]{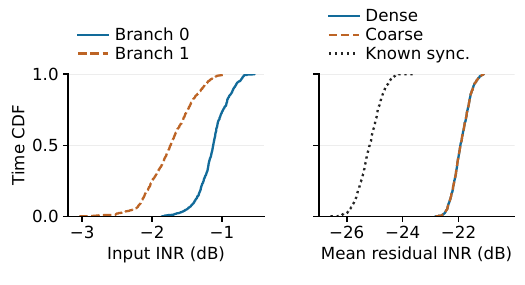}
\par\vspace{-3pt}
\caption{Input and mean residual INR CDFs over 716 equally weighted
Starlink snapshots. Dense and coarse residual curves nearly coincide.}
\label{fig:conference_timeseries}
\vspace{-3pt}
\end{figure}

\section{Long-Term Starlink Constellation Evaluation}
\label{sec:long_term_evaluation}

\begin{table}[t]
\caption{Starlink snapshot parameters (rounded where shown).}
\label{tab:starlink_parameters}
\centering\footnotesize
\setlength{\tabcolsep}{4pt}
\begin{tabular}{@{}ll@{}}
\toprule
Parameter & Value\\
\midrule
Ground latitude / longitude & $39.957^\circ$ / $-105.160^\circ$\\
Ground height / receive gain & 1660 m / 0 dBi\\
Carrier frequency / EIRP & 10 GHz / 50 dBm\\
Temperature / noise figure & 290 K / 5 dB\\
Noise bandwidth / noise power & 244.14 kHz / $-115.10$ dBm\\
Sample rate / samples per symbol & 1.171875 MHz / $S=6$\\
Record length / FIR length & $N=2916$ / 192 taps\\
Delay / Doppler search range & $\pm0.30$ ms / $\pm125$ kHz\\
Training trials / test blocks per trial & 120 / 32\\
\bottomrule
\end{tabular}
\end{table}

A separate 24-hour experiment samples 720 orbital snapshots at 120-s
intervals using the scalar LoS link budget in
Table~\ref{tab:starlink_parameters}. Among satellites with elevation
at least $60^\circ$, branch~0 is nearest. Branch~1 is selected uniformly
with a fixed seed from partners satisfying the tabulated relative delay
and Doppler limits. This pairing rule retains 716 snapshots and keeps
the dense FFT baseline tractable. Both methods search the same bounded
region.

Bandlimited thermal noise is modeled in each branch-centered ideal
receive passband, which preserves the full RRC spectrum. Absolute Doppler
is retained. Carrier centers are not supplied to the estimator.
Output-noise normalization accounts for temporal correlation.

Both methods use matched final FIR refinement with at most 30 iterations
in a $\pm1.2$-sample, $\pm0.6$-bin box around the initial estimate,
clipped to the search limits. Per-snapshot residuals average the tabulated
training trials and independent test blocks in linear power.

In Fig.~\ref{fig:conference_timeseries}, branch input INRs range over
$[-1.86,-0.54]$ and $[-3.04,-0.97]$~dB. Both receivers achieve a
median per-snapshot mean residual of $-21.94$~dB and a 95th percentile
of $-21.37$~dB, with no observed wrong acquisition in 85,920 training
trials. The known-synchronization median is $-25.18$~dB, revealing a
remaining alignment penalty. These CDFs summarize per-snapshot means,
not pooled trials. Adjacent snapshots are temporally correlated.
Different noise bandwidth and record length preclude testing the
fixed-geometry $2/1222$ plateau here.

%% file: paper_sector/conference_short/conclusion.tex
\section{Conclusion}
The mean residual INR approximation connects acquisition, synchronization,
and finite-training losses, predicting a $2/N_v$ plateau under stated
assumptions and quantifying imbalance and bias effects. With matched
refinement, coarse search approaches dense-search nulling after reliable
acquisition while reducing measured search time by $6.7\times$ at a
roughly 0.7-dB acquisition penalty. Starlink snapshots support comparable
nulling under changing geometry. Desired-user synchronization,
signal-preserving combining, and equalization where needed remain
future work.